\documentclass[9pt]{IEEEtran}

\usepackage[cmex10]{amsmath}
\usepackage{amsfonts}
\usepackage[font=footnotesize]{caption}
\usepackage{cite}
\usepackage{amsthm}
\usepackage{amssymb}
\usepackage{circuitikz}
\usepackage{tikz}
\usepackage{subfig}    
\usepackage{float}
\usetikzlibrary{hobby}
\usepackage[flushleft]{threeparttable}
\usepackage[normalem]{ulem}
\usepackage{pgfplots}
\usepackage{tabularx} 
\usepackage{diagbox}
\usepackage[flushleft]{threeparttable}
\usepackage{pifont}
\usepackage[colorinlistoftodos,prependcaption,textsize=small]{todonotes}
\usepackage{tablefootnote}
\usepackage{regexpatch}
\makeatletter
\xpatchcmd{\@todo}{\setkeys{todonotes}{#1}}{\setkeys{todonotes}{inline,#1}}{}{}
\makeatother

\newcommand{\J}{\mathbf{J}}

\newcommand{\R}{\mathbf{R}}
\newcommand{\X}{\mathbf{X}}

\newcommand{\Z}{\mathbf{Z}}

\begin{document}
\title{Shape Synthesis and 3D Ceramic Printing of Non-canonical MIMO Dielectric Resonator Antennas}
\author{Binbin Yang,~\IEEEmembership{Senior Member,~IEEE}, Jaewoo Kim, Trupti Bellundagi, and Jacob J. Adams,~\IEEEmembership{Senior Member,~IEEE}
\thanks{This work was supported in part by the US National Science Foundation under grant 2138741.}
\thanks{B. Yang and T. Bellundagi are with the Department of Electrical and Computer Engineering, North Carolina Agricultural and Technical State University, Greensboro, NC, USA (email: byang1@ncat.edu).}
\thanks{J. J. Adams and J. Kim are with the Department of Electrical and Computer Engineering, North Carolina State University, Raleigh, NC, USA (e-mail: jacob.adams@ncsu.edu).}

}


\maketitle

\begin{abstract}
In this paper, we report a shape synthesis method for multi-mode dielectric resonator antennas (DRA) using characteristic mode theory (CMT) and a binary genetic algorithm (BGA). By including the antenna's characteristic modal responses (resonance frequencies and quality factors) in the cost function, the shape synthesis process is conducted without including excitation feeds. Through the optimization procedure, a non-canonical dielectric body is formed from tetrahedral elements to support the required modal properties.  As a demonstration of the proposed design approach, two three-mode MIMO DRAs are synthesized from both a rectangular and a cylindrical volume to operate at 2.45 GHz. The synthesized MIMO DRA's complex shape (based on rectangle) is then fabricated using Nanoparticle jetted zirconia. A combination of probe and slot feeds are employed to excite the desired modes. Due to the orthogonality of the characteristic modes and the careful design of the feeding network, isolation $>20$ dB is achieved between all ports.
\end{abstract}


\section{Introduction}

\IEEEPARstart{D}{ielectric} resonator antennas (DRAs) have drawn significant attention from researchers and engineers worldwide since their initial investigation by Long \cite{long1983}. DRAs exhibit high radiation efficiency due to lack of conductive loss, and the DRA remains highly efficient even at millimeter wave frequency range~\cite{lai2008comparison} compared to conductive antennas, such as the microstrip patch. DRAs also feature compact size due to their high dielectric constant.  In addition, one of the most appealing features of the DRA is the design freedom available with a 3-dimensional resonator made from a selectable permittivity.

Owing to this flexibility, DRA designs have been reported in literature covering various applications, such as broadband designs \cite{chair2004wideband,huang2007compact}, multi-port applications \cite{yan2011design,abdalrazik2017three}, beamforming arrays \cite{chow1995cylindrical,su2016linearly}, and millimeter wave applications \cite{zhang2019mimo,nor2016rectangular}. However, most DRA designs are based on canonical shapes, such as rectangular, cylindrical and spherical geometries \cite{Leung1993Theory,abdalrazik2017three,fang2014theory}, for which analytical design formulas and conventional manufacturing techniques are most suitable. While designs based on canonical geometries are simple and straightforward, they limit design and research possibilities. Given the recent advancements in additive manufacturing of high permittivity materials \cite{roper2014additive,kadvera2022wide,lou2020design,oh2019microwave,Oh2023}, it has become feasible to explore more complex 3D  DRA designs.

Indeed, several researchers have explored DRA designs using non-canonical geometries ~\cite{xia20193D,trinh2016wideband,alroughani2020shape}. In \cite{xia20193D}, a broadband DRA with a simple stepped-radial distribution having four permittivity values is printed using an additive manufacturing process. However, details on the design approach are not given, though it appears that the four steps were selected through a brief optimization. More complex optimizations and structures have also been considered.  In \cite{trinh2016wideband}, a broadband DRA is optimized by division into a 2D array of pillars with various heights that are subject to optimization with a genetic algorithm (GA) and later machined to height.  However, only varying heights of 64 dielectric bars are considered, limiting the range of valid structures. Finally, in the approach most similar to this work, \cite{alroughani2020shape} conducts synthesis of two-port DRA based on characteristic modes.  The authors discretize the DRA into a coarse grid of cubes and perform a GA optimization while placing some constraints on the modal patterns and feed locations.

In this work, we employ a novel method for DRA non-canonical synthesis derived from our work on microstrip antenna shape-first, feed-next synthesis~\cite{yang2016systematic,yang2019shape}. The method optimizes the fundamental properties of the characteristic modes of the structure using a binary genetic algorithm. We assess the fundamental properties of each characteristic mode including their resonant frequencies and Q factors and use these to guide the optimization, followed by a feed design step to independently excite these modes. We use a volume integral equation (VIE) approach with tetrahedral discretization, and the optimization degrees of freedom coincide with these tetrahedra to resolve complex geometries with fine geometric features, both internal and external.  For example, our demonstration cases use more than 10 times the number of degrees of freedom used in~\cite{alroughani2020shape}. Finally, a state-of-the-art ceramic printing process is used to realize these fine features.

Section II presents the background and optimization framework for the proposed synthesis process. In Section III, the process is used to synthesize two three-port MIMO DRAs composed of zirconia.  Finally, Section IV describes the feed design, fabrication, and measured characteristics of the device.


\section{Shape Synthesis of MIMO DRAs}
\label{section:synthesis}


Characteristic mode theory (CMT) provides a systematic and efficient method for analyzing the behavior of complex antenna structures~\cite{harrington1971theory}. By decomposing the antenna into its characteristic modes, a design can assess and tune the properties of the dominant modes.  This enables systematic optimization techniques that relate changes in the antenna geometry with changes in the underlying eigenmodes in a more direct way than when observing the aggregate modal response. Moreover, CMT optimization can be conducted without prior knowledge of a feed arrangement~\cite{ethier2014antenna,yang2019shape}.  The utility of CMT for metallic antenna synthesis has been reported in various works \cite{ethier2014antenna,yang2016systematic,yang2019shape,filtering_antenna_synthesis,ESA_synthesis}.  However, its application to DRA synthesis has been limited. Here we investigate the application of CMT to DRA synthesis
using a shape-conformal discretization and physical parameter driven optimization framework, and demonstrate its application to the design of non-canonical MIMO DRAs.

\subsection{VIE-based Characteristic Mode Analysis}
To decompose the antenna into its characteristic current modes, we solve the following eigenvalue equation \cite{harrington1971theory}:
\begin{equation}
\mathrm{\X\J_n=\lambda_n \R \J_n},
    \label{eq:CMT}
\end{equation}
where $\mathrm{\X}$ and $\mathrm{\R}$ are the imaginary and real parts of the method of moments (MoM) $\mathrm{\Z}$ matrix, and $\mathrm{\lambda_n}$ and $\mathrm{\J_n}$ are the eigenvalue and eigencurrent of the $n-$th mode. The MoM $\mathrm{\Z}$ matrix can be based on the surface integral equation (SIE) or volume integral equation (VIE). As pointed out in \cite{alroughani2013appraisal}, the VIE, though more computationally expensive than the SIE, avoids the emergence of non-physical modes in the characteristic mode analysis of finite dielectric bodies. Furthermore, the equivalent polarization currents calculated from the VIE can be directly used to calculate Q factors following the source formulation as demonstrated in \cite{yang2017quality,yang2021fundamental}, whereas it's unclear if the equivalent surface currents in SIE can be used to accurately calculate the antenna stored energy and Q factor. Moroever, the VIE based on tetrahedral discretization allows natural synthesis of internal fine features, whereas the SIE based on surface discretization has very small efficiency advantage when the object has a large surface area relative to its volume, as expected with a complex internal geometry. We therefore employ the VIE for our CM analysis and shape synthesis.

\textcolor{black}{The VIE MoM formulation is numerically implemented in our in-house Matlab CMA solver with 64-point Gaussian quadrature for the integral evaluation, and its accuracy has been validated in some of our earlier work \cite{yang2017quality,yang2021fundamental}. The same code is used for the numerical study in this work.} 

\begin{figure}
\centering 

{\includegraphics[width=0.7\linewidth]{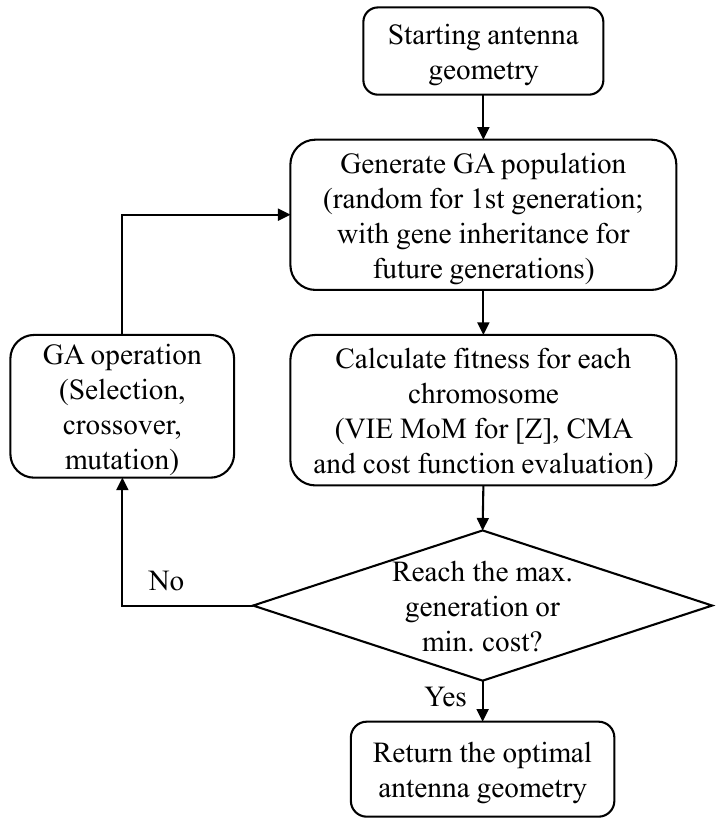}%
}
\caption{The MIMO DRA synthesis process based on the VIE-based CMA and binary genetic algorithm.}
\label{fig:DRA_optimization_process}
\end{figure}

\subsection{Synthesis Process and Cost Function}
The synthesis process used here is adopted from the shape-first, feed-next antenna design method for planar metallic antennas introduced in \cite{yang2019shape}. \textcolor{black}{The detailed workflow of the optimization process is illustrated in Figure \ref{fig:DRA_optimization_process}.}
Here we use the tetrahedral discretization of the dielectric structure as a basis for the fine features of the design.  Assigning a binary "gene" to each tetrahedra to represents its assignment as air or the selected dielectric medium, we employ a genetic algorithm to iteratively evaluate and evolve through thousands of antenna designs. Because the tetrahedra represent a fixed basis, the matrix fill step needs to be computed only once~\cite{yang2019shape} and we are able to efficiently evaluate new combinations. 

For the MIMO DRA synthesis problem studied here, the goal is to search for an antenna geometry with multiple modes resonating at the same frequency with the lowest possible Q factors (\textit{i.e.}, for maximum bandwidth). As shown in \cite{yang2017quality,yang2021fundamental}, characteristic modal Q factors $\mathrm{Q_n}$ for each DRA can be calculated from their characteristic modal current distribution. While minimizing the Q factor maximizes the potential bandwidth of the mode, we also seek modes that are near resonance, which makes them easier to excite in isolation, so we also compute the characteristic modal significance (MS) of each mode defined by $\mathrm{MS_n=\frac{1}{|1+j\lambda_n|}}$. The MS reaches a maximum of 1 at self resonance ($\lambda_n=0$). Finally, it is assumed that the antenna must conform to a pre-defined volumetric superstructure that contains all of the possible tetrahedra.  To facilitate the development of a compact MIMO DRA, the cost function for the shape synthesis procedure is defined from characteristic modal parameters as: 
\begin{equation}
    \mathrm{cost=\sum_{n=1}^N C_n}+w_2\mathrm{V},
    \label{eq:cost_function}
\end{equation}
where $\mathrm{C_n=}w_1\mathrm{(1-MS_n)+Q_n}$ represents the cost contributed by the $n-$th mode.  $\mathrm{V}$ is the volume of the studied geometry normalized to that of the superstructure.  Minimizing this term removes unnecessary material, reducing size and weight. Weighting coefficients $w_1$ and $w_2$ are selected depending on the problem. 
\begin{figure}
\centering 
{\subfloat[]{\includegraphics[width=0.49\linewidth]{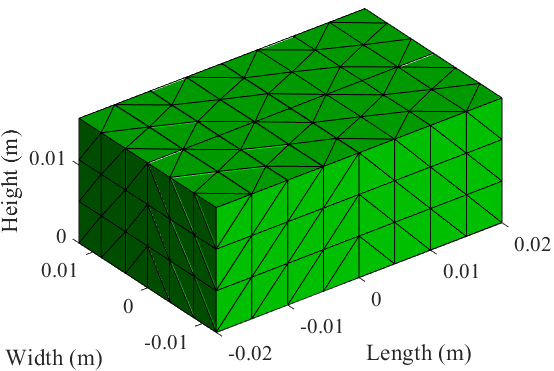}%
\label{fig:a}}}
\hfil
{\subfloat[]{\includegraphics[width=0.49\linewidth]{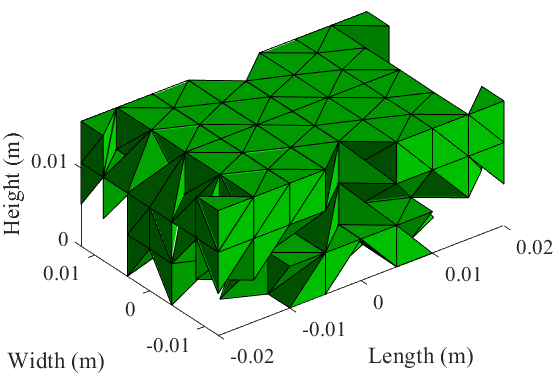}%
\label{fig:b}}}
{\subfloat[]{\includegraphics[width=0.75\linewidth]{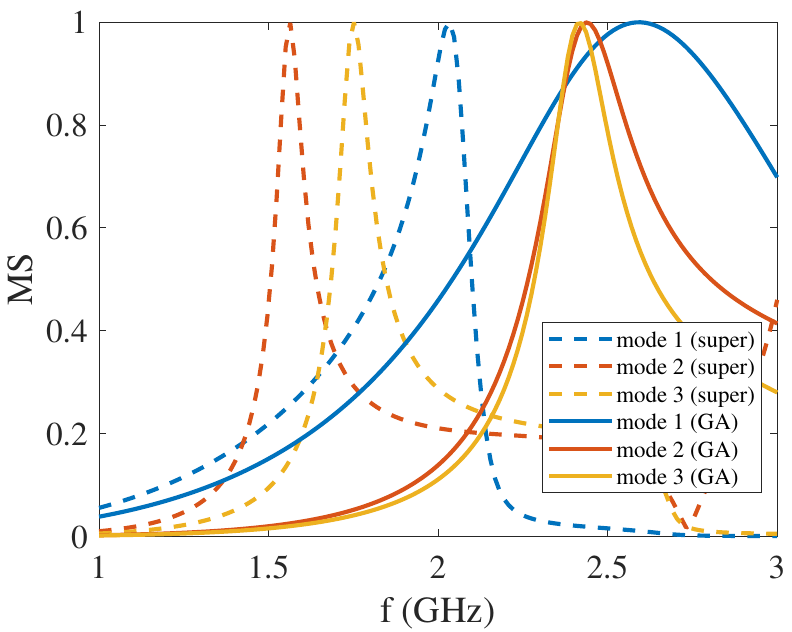}%
}}
\caption{(a) The complete rectangular DRA super geometry (super), (b) the optimized rectangular DRA geometry (GA), (c) the modal significance before (dashed lines) and after (solid lines) shape synthesis. (relative permittivity is 23)}
\label{fig:design_example}
\end{figure}

\begin{figure}
\centering%
\subfloat[]{%
\centering
\includegraphics[width=0.5\linewidth]{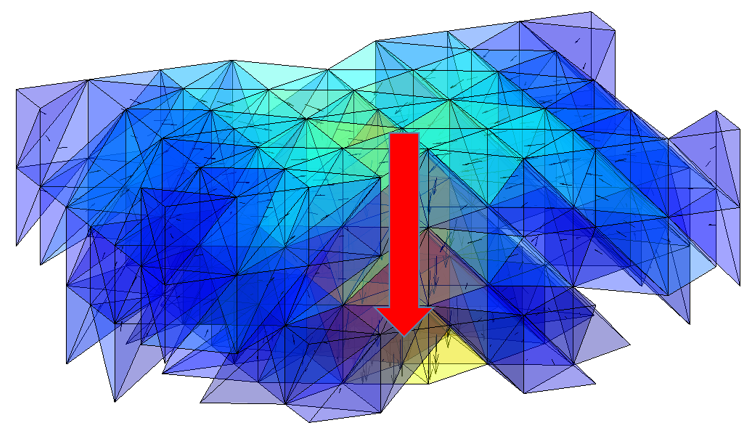}
\label{fig:current_mode1}
}%
\subfloat[]{%
\centering
\includegraphics[width=0.5\linewidth]{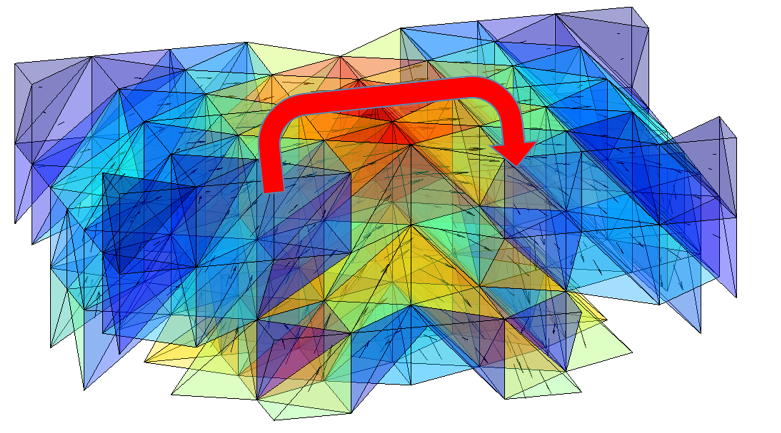}
\label{fig:current_mode2}
}%
\\
\subfloat[]{%
\centering
\includegraphics[width=0.47\linewidth]{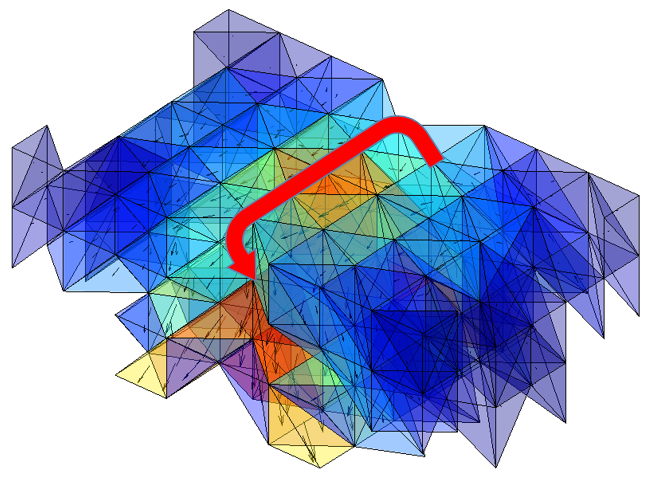}
\label{fig:current_mode3}
}%
\caption{\textcolor{black}{The characteristic modal polarization currents of the synthesized rectangular DRA at 2.45 GHz: (a) mode 1; (b) mode 2; and (c) mode 3.}}
\label{fig:modal current}
\end{figure}



\section{Shape Synthesis of Three-port MIMO DRAs}
As demonstration of the flexibility of the proposed synthesis process for MIMO DRAs, two MIMO DRAs are synthesized within a rectangular and a cylindrical volume at 2.45 GHz. Both optimized to support three decoupled ports at the design frequency. To experimentally verify the design approach, the synthesized rectangular DRA is prototyped and experimentally characterized.

\subsection{DRA Synthesis on A Rectangular Block}
The first shape synthesis example is based on a rectangular superstructure DRA with a dimension of $\mathrm{40 \times 25 \times 16 mm^3}$, as shown in Fig. \ref{fig:design_example} (a). The optimization is targeted at finding a 3-mode DRA operating at 2.45 GHz. The material assumed for this DRA design is zirconia, with a dielectric constant of $\epsilon_r=23$, and a loss tangent of $\tan\delta=0.0015$~\cite{oh2019microwave}. The entire DRA is placed above an infinite ground plane which facilitates antenna excitation after the shape synthesis.

It is important to note that the modal resonant frequencies of any sub-structure DRA will always be greater than or equal to the corresponding mode of the super-structure~\cite{yang2021fundamental}. In other words, since three modes are desired at 2.45 GHz in this case, the superstructure dimensions must be selected so that at least three modes of the entire dielectric body are resonant at a frequency below 2.45 GHz.  Any shape changes that reduce the volume of the dielectric body can only raise the modal resonant frequencies. The rectangular DRA shown in Fig.~\ref{fig:design_example}(a) has modal resonance frequencies of 2.03 GHz, 1.56 GHz and 1.75 GHz. Since these are all below the design frequency of 2.45 GHz, we anticipate that the DRA shape synthesis procedure can shift each mode to the design frequency.

To synthesize the DRA that meets these requirements, the rectangular DRA is discretized into 720 tetrahedra with two plane symmetries enforced along the length and width direction. A total of 300 generations are searched, with a population of 40 antennas for each generation and a mutation rate of 0.15. The cost function in (\ref{eq:cost_function}) with $N=3$, $w_1=200$ and $w_2=20$ is applied to each generation to select the best fit designs. Figure \ref{fig:design_example}(b) shows the final synthesized DRA geometry and Figure \ref{fig:design_example}(c) compares the characteristic modal significance before and after shape synthesis. As shown in the solid lines in Figure \ref{fig:design_example}(c), the first three characteristic modes of the synthesized DRA are all resonant around the design frequency 2.45 GHz, whereas the modes of the complete rectangular DRA (dashed lines) are originally resonant at lower frequencies (2.03 GHz, 1.56 GHz and 1.75 GHz). Table \ref{tab_rect} compares the modal significances (MS) and Q factors for the first three modes before and after resonance at 2.45 GHz. Noticeably, all the MS values are now close to 1, which indicates self resonance, and the modal Q factors have been reduced for all the three modes, indicating a wider fractional bandwidth after the shape synthesis. The characteristic modal polarization currents of the first three modes on the synthesized DRA is given in Figure \ref{fig:modal current}. Due to the presence of the ground plane, the first modal current is similar to a monopole, and the 2nd and 3rd modes are similar to half vertical loops over ground in two orthogonal directions. The modal current distribution will facilitate feeding scheme designs as discussed later.

\begin{table}[t]
\caption{Comparison of Modal Significances and Modal Q Factors of the rectangular DRA Before and After Shape Sythesis at 2.45 GHz}
\renewcommand{\arraystretch}{1.35}
\begin{center}
\begin{tabular}{c|c|c|c|c}
\hline\hline
\textbf{Modes} &  \textbf{{MS (Before)}}& \textbf{{MS (After)}} &\textbf{{Q (Before)}}& \textbf{{Q (After)}} \\
\hline
Mode 1& 0.02 & 0.94 & 429 & 2.98\\
Mode 2& 0.18 & 0.99 & 18.8 & 9.20\\
Mode 3& 0.20 &  0.95 & 16.3 & 12.7\\
\hline

\end{tabular}
\label{tab_rect}
\end{center}
\end{table}

\begin{figure}
\centering 
{\subfloat[]{\includegraphics[width=0.49\linewidth]{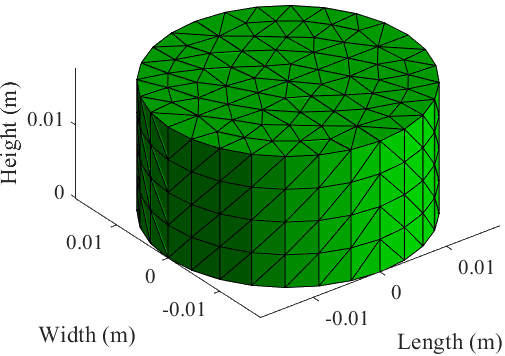}}%
\label{fig:a}}
\hfil
{\subfloat[]{\includegraphics[width=0.49\linewidth]{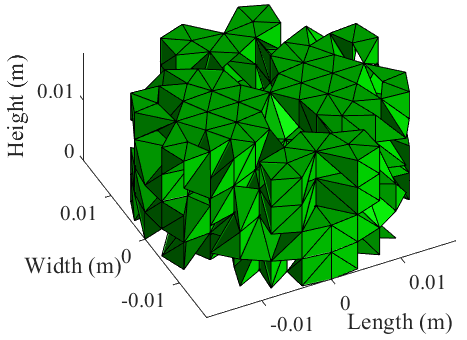}}%
\label{fig:b}}
{\subfloat[]{\includegraphics[width=0.75\linewidth]{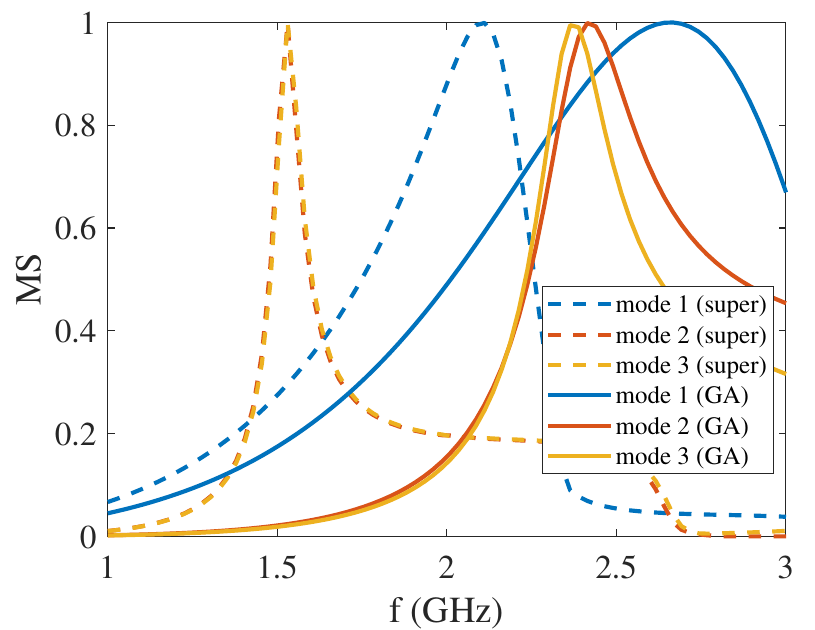}%
}}
\caption{(a) The complete cylindrical DRA super geometry (super), (b) the optimized cylindrical DRA geometry (GA), (c) the modal significance before (dashed lines) and after (solid lines) shape synthesis. (relative permittivity is 23)}
\label{fig:design_example_cylinder}
\end{figure}



\subsection{DRA Synthesis on A Cylindrical Block}
The advantage of the proposed synthesis approach is that it can synthesize optimal DRA designs for arbitrary geometric constraints. 
 As another demonstration, a second shape synthesis is conducted on a cylindrical superstructure DRA with a radius of 18 mm and a height of 18 mm, as shown in Fig. \ref{fig:design_example_cylinder}(a). The same zirconia material and cost function as the previous case are adopted, and the DRA is placed above an infinite ground plane with two planes of symmetries being enforced. Prior to optimization, the modal resonance frequencies of the full cylindrical DRA are measured as 2.10 GHz, 1.53 GHz and 1.53 GHz (degenerate), which should allow optimization to the higher target frequency of 2.45 GHz.

\begin{figure}
\centering%
\subfloat[]{%
\centering
\includegraphics[width=0.4\linewidth]{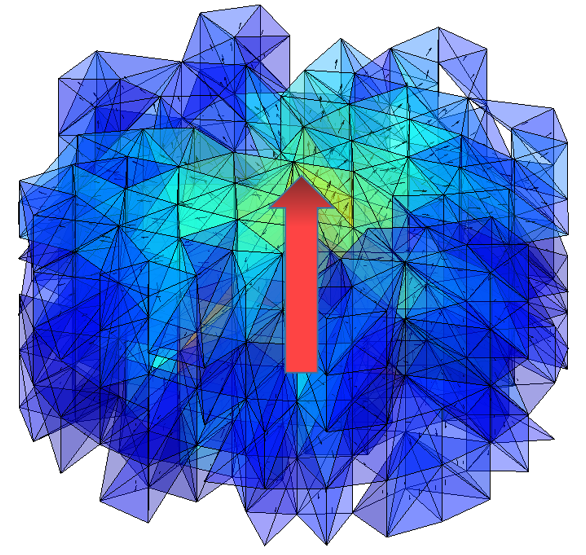}
\label{fig:current_mode1}
}%
\subfloat[]{%
\centering
\includegraphics[width=0.38\linewidth]{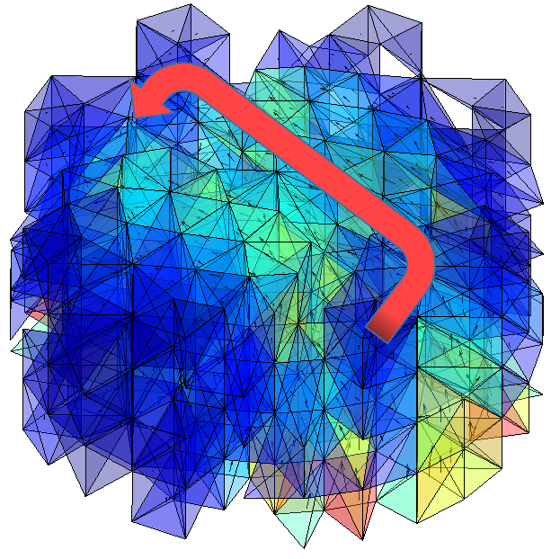}
\label{fig:current_mode2}
}%
\\
\subfloat[]{%
\centering
\includegraphics[width=0.40\linewidth]{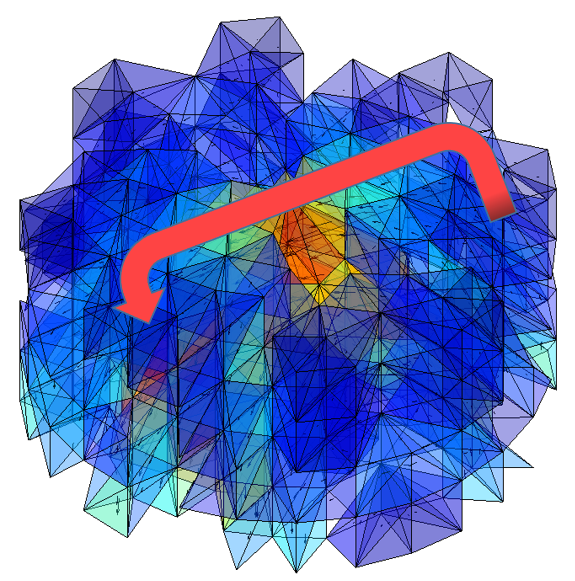}
\label{fig:current_mode3}
}%
\caption{\textcolor{black}{The characteristic modal polarization currents of the synthesized cylindrical DRA at 2.45 GHz: (a) mode 1; (b) mode 2; and (c) mode 3.}}
\label{fig:modal current cylinder}
\end{figure}

Following the synthesis procedure, Figure \ref{fig:design_example_cylinder}(b) shows the optimized DRA geometry that supports three modes around 2.45 GHz, and Figure \ref{fig:design_example_cylinder}(c) compares the modal significance of the first three CM modes on the original cylindrical DRA and the optimized cylindrical DRA. Table \ref{tab_cylinder} compares the modal significances (MS) and Q factors for the first three modes before and after resonance at 2.45 GHz. Again, shape synthesis brings all the three modes from 2.10 GHz, 1.53 GHz and 1.53 GHz respectively close to the design frequency of 2.45 GHz, and the modal Q factors have been reduced for all the three modes, indicating an increase in fractional bandwidth after the shape synthesis. Figure \ref{fig:modal current cylinder} shows the first three modal currents on the synthesized cylindrical DRA. Similar current distributions are observed as the rectangular case due to the similar form factor of the two designs.

\begin{table}[t]
\caption{Comparison of Modal Significances and Modal Q Factors of the Cylindrical DRA Before and After Optimization at 2.45 GHz}
\renewcommand{\arraystretch}{1.35}
\begin{center}
\begin{tabular}{c|c|c|c|c}
\hline\hline
\textbf{Modes} &  \textbf{{MS (Before)}}& \textbf{{MS (After)}} &\textbf{{Q (Before)}}& \textbf{{Q (After)}} \\
\hline
Mode 1& 0.06 & 0.91 & 326,000\tablefootnote{Note that the Q factor before shape synthesis can be very large given that the modal Q factor generally increases dramatically after passing the resonance frequency \cite{yang2017quality}.} & 2.50\\
Mode 2& 0.17 & 0.98 & 20.6 & 8.46\\
Mode 3& 0.17 &  0.84 & 20.6 & 11.1\\
\hline
\end{tabular}
\label{tab_cylinder}
\end{center}
\end{table}

\section{Excitation Schemes for the Synthesized Rectangular MIMO DRA}
\label{section:feed}
\begin{figure}
\centering 

{\includegraphics[width=0.55\linewidth]{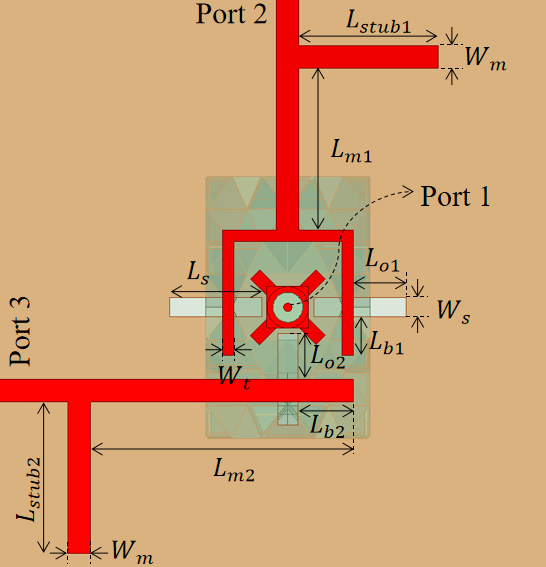}%
\label{fig:a}}
\caption{The bottom view of the feeding network. Port 1 is excited by SMA probe. Port 2 and 3 are excited by microstrip lines coupled with slots.}
\label{fig:feeding_network}
\end{figure}

\begin{figure}[t]
\centering 
{\subfloat[]{\includegraphics[width=0.47\linewidth]{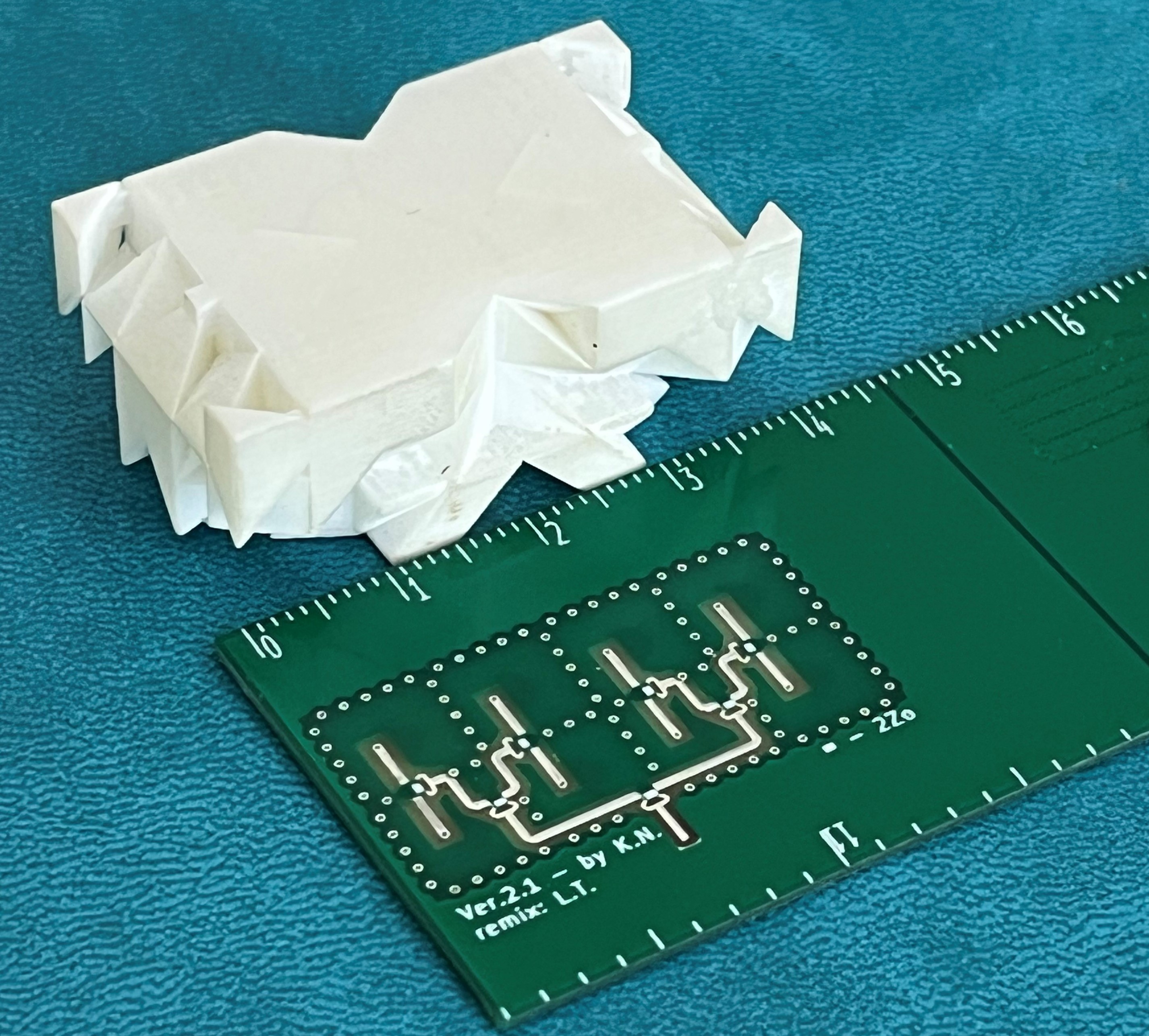}%
\label{fig:a}}}
{\subfloat[]{\includegraphics[width=0.47\linewidth]{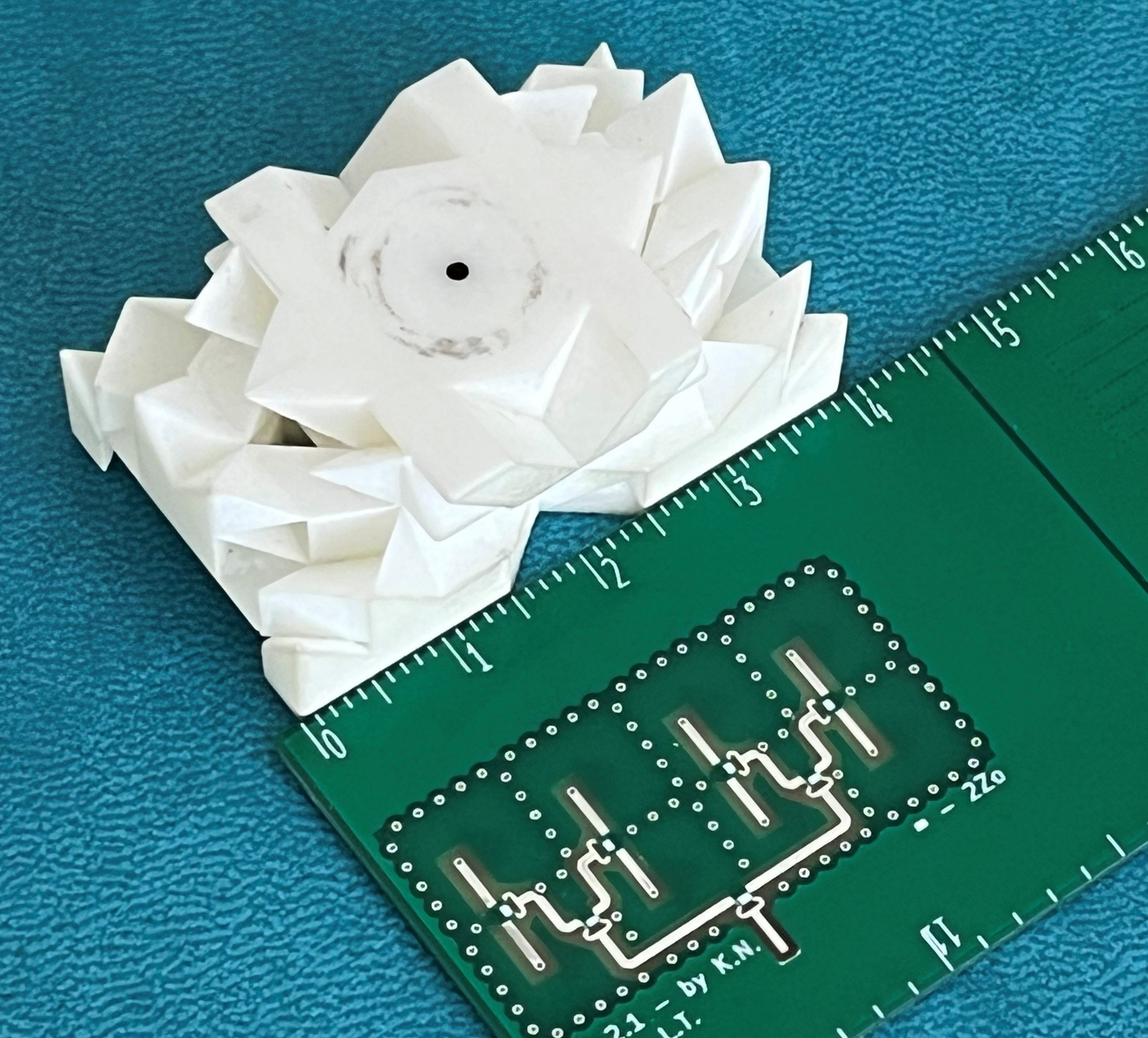}%
\label{fig:a}}}
{\subfloat[]{\includegraphics[width=0.47\linewidth]{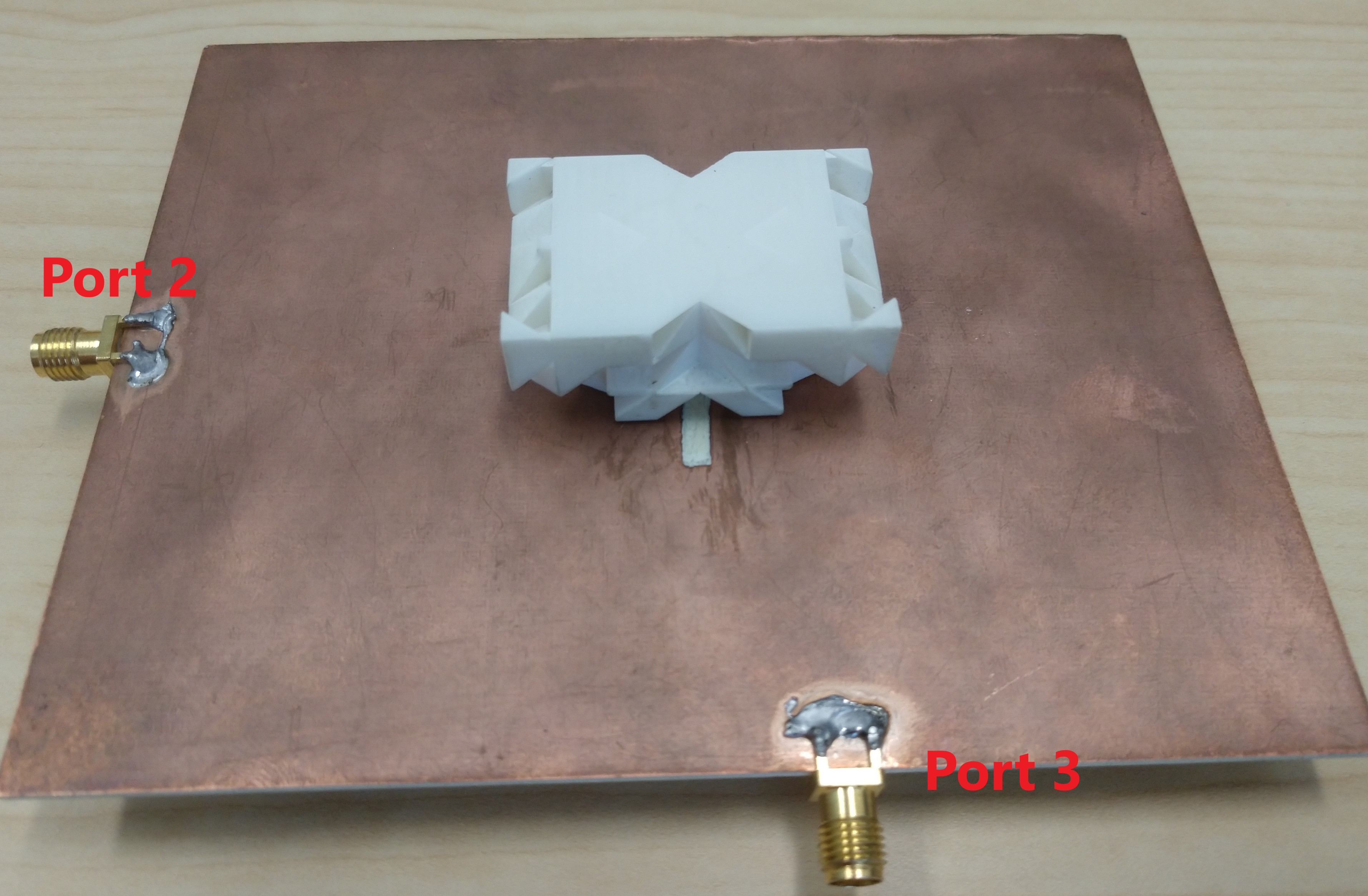}%
\label{fig:a}}}
{\subfloat[]{\includegraphics[width=0.47\linewidth]{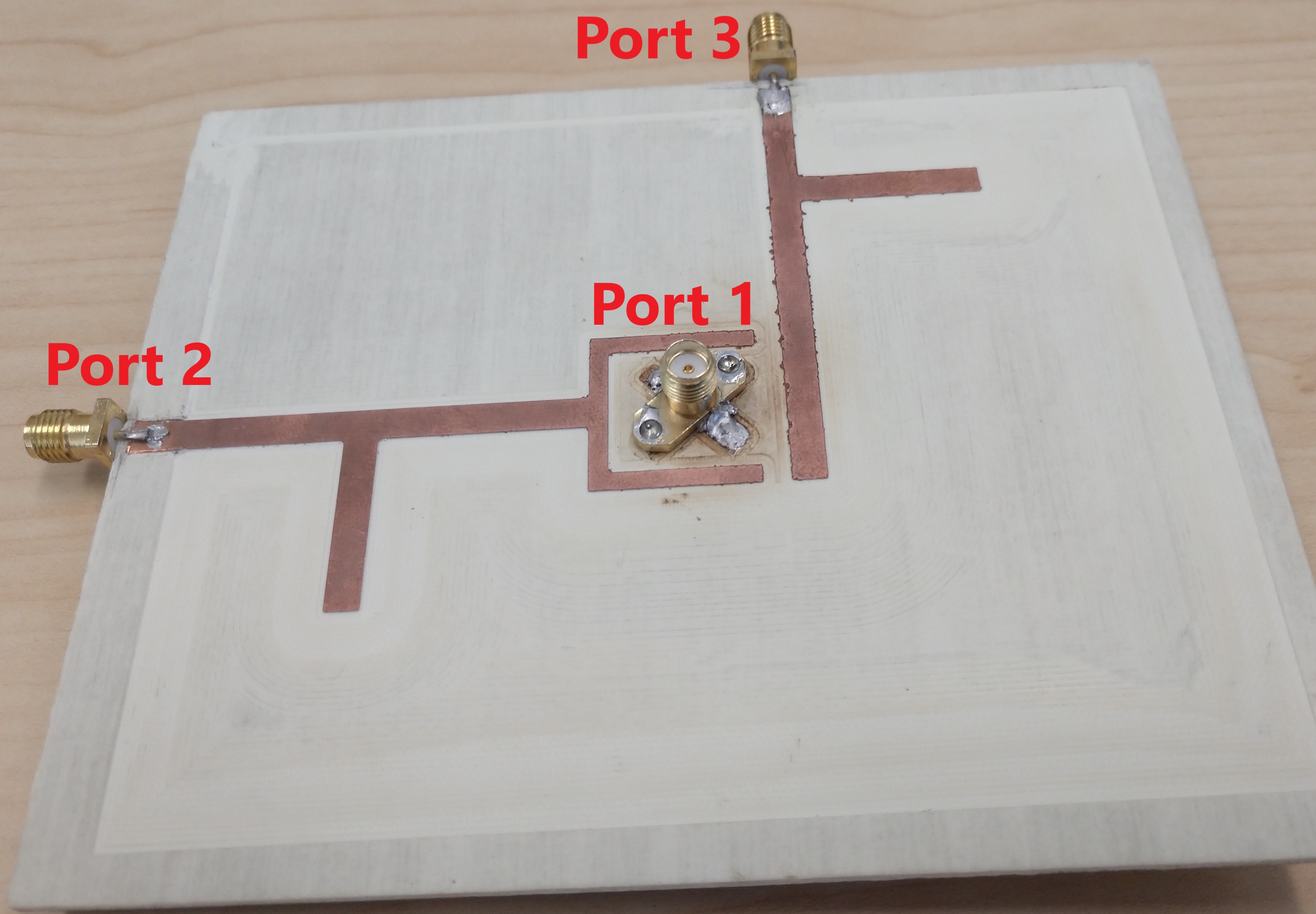}%
}}

{\subfloat[]{\includegraphics[width=0.8\linewidth]{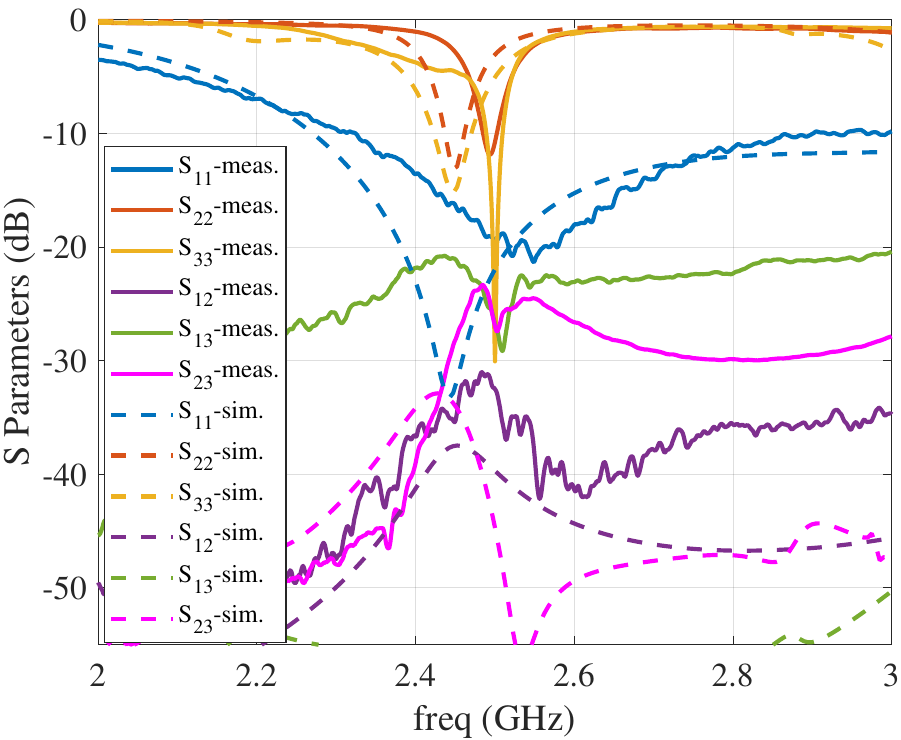}%
}}
\caption{(a) the DRA prototype top-side view; (b) the DRA prototype bottom-side view; (c) the MIMO DRA placed on its feeding network; (d) the bottom view of the prototyped feeding structure; and (e) the measured S parameters (solid lines) in comparison with simulation results (dashed lines).}
\label{fig:prototype}
\end{figure}

After obtaining the DRA geometry that supports the required modal resonant frequencies through shape synthesis, the next step is to design the feeding network that excites these modes and achieves both good input impedance matching and high isolation. We here take the rectangular DRA as the demonstration example. Based on the modal polarization current distributions in Figure \ref{fig:modal current} and partly inspired by the work in \cite{fang2014theory}, a feeding network is proposed as shown in Figure \ref{fig:feeding_network}. Probe feeds were initially explored as an simple excitation scheme for all the modes, but they were found to exhibit significant coupling between the ports. To best excite the three modes with high isolation, a combination of slot and probe feeds are employed. In observation of the vertical current distribution for mode 1, port 1 (mode 1) is excited using a vertical probe. For port 2 (mode 2), with a polarization current distribution of a half loop, dual slots are used to couple the energy from the feed line to the modal magnetic fields. A power divider is used as the microstrip feeding structure. Port 3 (mode 3) is excited similarly, but due to limited space, a single slot is employed to couple the energy from microstrip feed line to the magnetic fields. The microstrip substrate is Rogers 4350 with the height of $h=1.52$ mm, relative permittivity $\epsilon_r=3.48$, and loss tangent $\tan\delta =0.0037$. In reference to Fig.~\ref{fig:feeding_network}, the parameters for the microstrip feed network are given in Table \ref{table:I}. 
The two microstrip stub lines $L_{stub1}$ and $L_{stub2}$ are used for impedance matching. 
The simulated response for the three-port MIMO DRA is shown in Figure \ref{fig:prototype}(e) as dashed lines. Both good impedance matching and high isolation is achieved at 2.45 GHz.

\begin{figure}[!t]
\centering 
{\subfloat[]{\includegraphics[width=0.4\linewidth]{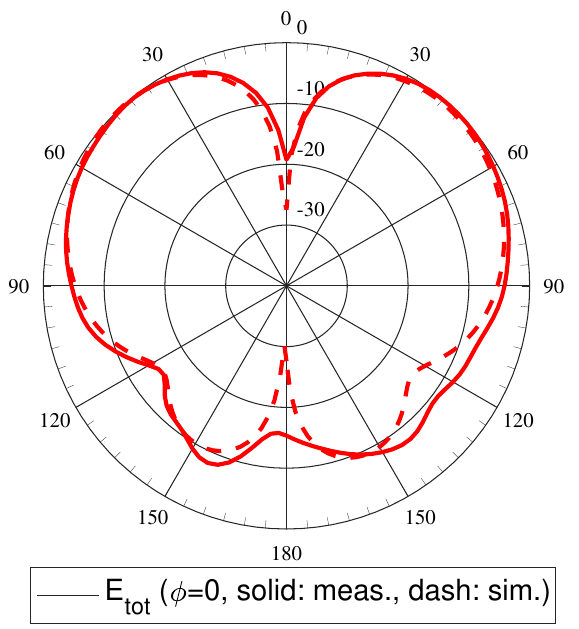}%
}}
{\subfloat[]{\includegraphics[width=0.4\linewidth]{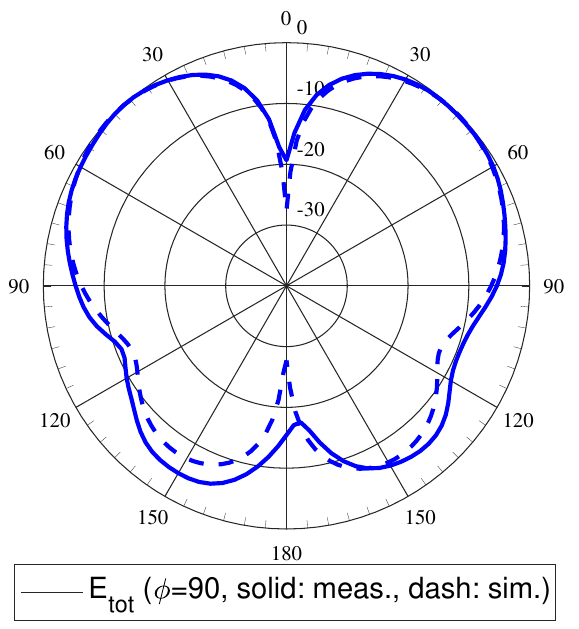}%
\label{fig:b}}}
{\subfloat[]{\includegraphics[width=0.4\linewidth]{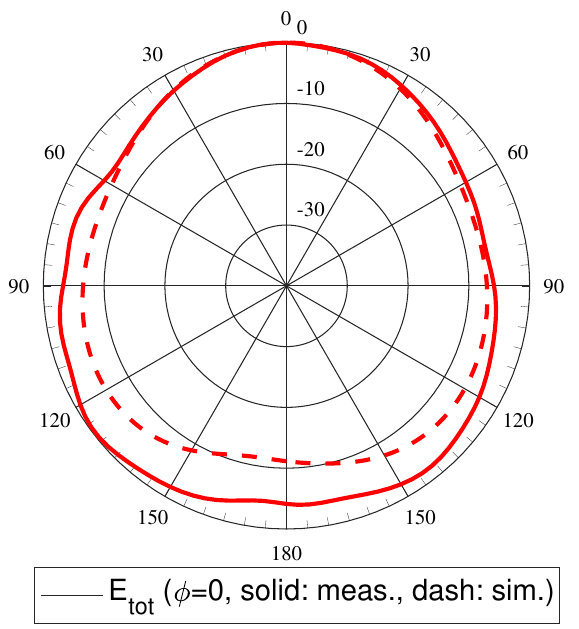}%
}}
{\subfloat[]{\includegraphics[width=0.4\linewidth]{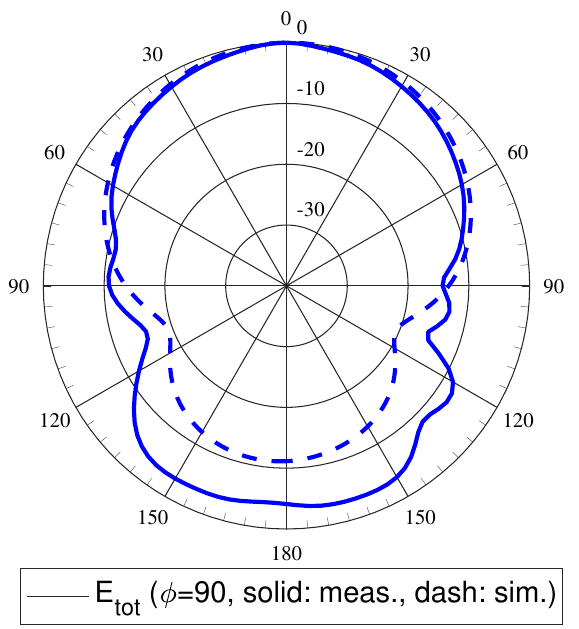}%
\label{fig:a}}}
{\subfloat[]{\includegraphics[width=0.4\linewidth]{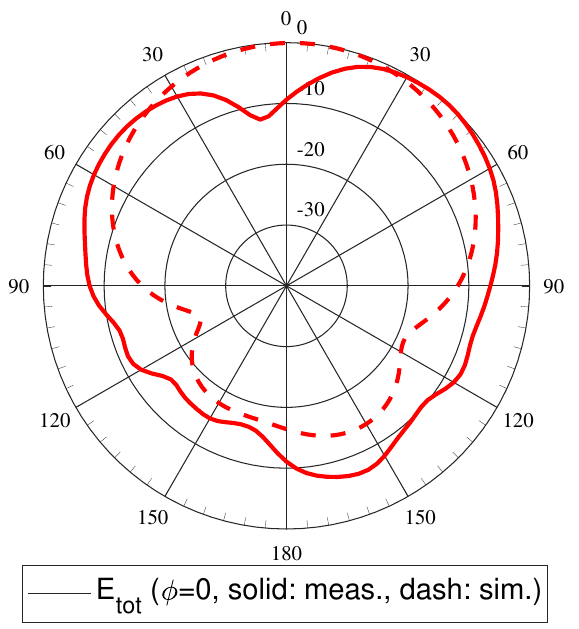}%
\label{fig:b}}}
{\subfloat[]{\includegraphics[width=0.4\linewidth]{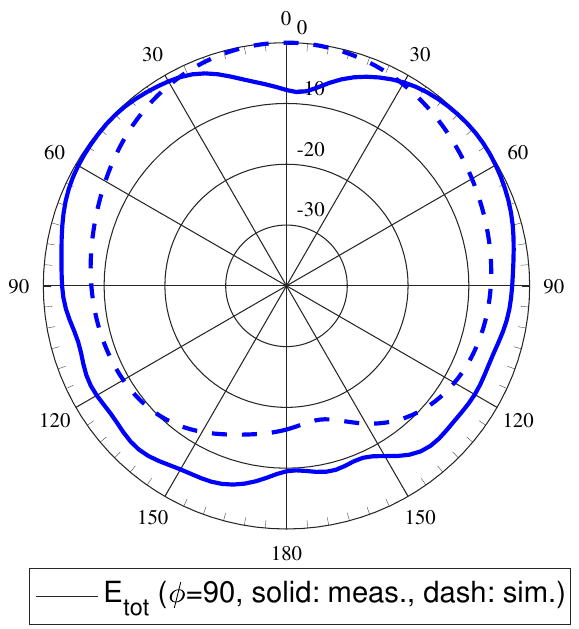}%
}}
\caption{(a) and (b) are the normalized radiation pattern of port 1; (c) and (d) are the normalized radiation patterns of port 2; (e) and (f) are the normalized radiation patterns of port 3. Solid lines are the measurement data and dashed lines are the simulation data. Red curves show the pattern at $\phi=0$, and the blue lines are the pattern at $\phi=90$ degree.}
\label{fig:farfield}
\end{figure}

\begin{table}[h]
\renewcommand{\arraystretch}{1.35}
\caption{Dimension Parameters of the Microstrip Feed Network (in mm)} 
\centering 
\begin{tabular}{c | c | c | c | c | c} 
\hline\hline 
Parameter    & Value & Parameter & Value & Parameter & Value\\
\hline 
$L_{m1}$  & 24.71 & $L_{b1}$ & 5.78 & $L_s$ & 14\\
$L_{m2}$  & 40.09 & $L_{b2}$ & 8.50 & $W_m$ & 3.45\\ 
$L_{stub1}$   & 21.20 & $L_{o1}$ & 8 & $W_t$  & 1.72 \\
$L_{stub2}$   & 23.10 & $L_{o2}$ & 7 & $W_s$ & 3\\
\hline 
\end{tabular}
\label{table:I} 
\end{table}

\subsection{Prototyping and Measurement}
\label{section:prototyping}
The optimized antenna geometry in Figure \ref{fig:design_example} (b) was fabricated using 3D printed zirconia with the the XJet
Carmel 1400 3D printer. The printer uses a type of material jetting known as a nanoparticle jetting and can create full density zirconia with 20 micron resolution.  The printed and sintered part is shown in Figure \ref{fig:prototype} (a) and (b). Figure \ref{fig:prototype} (c) and (d) show the designed feeding network for the multi-port DRA. The measured 3-port S parameters are given as solid lines in Figure \ref{fig:prototype} (e), in comparison with the simulation results in dashed lines. A shift in operating frequency of about 2\% is observed from 2.45 to about 2.5 GHz, and the isolation is lower than simulated but is still good in the operating band, at about 25 dB.   The overlapping 10 dB is around 17 MHz, limited by the bandwidth of modes 2 and 3, which have a higher Q factor within the selected bounding volume.

The far field responses of the antenna were characterized in the SG-64 chamber at the Wireless Research Center of North Carolina. Figures \ref{fig:farfield} (a)-(b) show the total E field for port 1 at $\phi=0 ^\circ$ and $\phi=90^\circ$ plane cut respectively at the measured center frequency of 2.5 GHz. Figure \ref{fig:farfield} (c)-(d) and (e)-(f) show corresponding responses for port 2 and 3 respectively at 2.5 GHz. Overall, close agreement is observed for ports 1 and 2, with port 3 showing some unexpected reduction in directivity normal to the ground plane. However, as seen in the correlation calculations below, the pattern correlations remain low.  In addition, the system efficiency for port 1 is measured as high as 96\% at 2.52 GHz, while the system efficiency for port 2 and 3 are measured as 63\% at 2.48 GHz and 82\% at 2.51 GHz. 

\begin{figure}[t]
\centering 

{\includegraphics[width=0.65\linewidth]{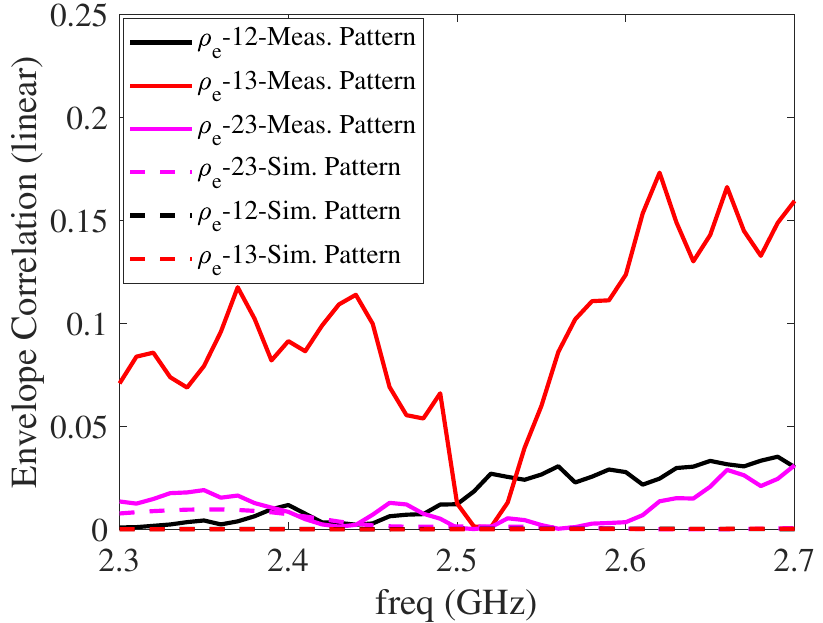}}%
\caption{The spatial envelope correlation coefficients between the three ports of the antenna calculated using antenna pattern (solid lines for measured data and dashed lines for simulation data).}
\label{fig:correlation}
\end{figure}

The envelope correlation coefficients between different ports are calculated based on the simulated and measured 3D field patterns \cite{blanch2003exact}, as shown in Figure \ref{fig:correlation}. The simulation result are given as dashed lines and the measured results are in solid lines. Overall, the correlation coefficients based on simulation results are close to 0 across the examined frequency range, indicating near-orthogonality between the three modes over a wide frequency range. The correlation data in the measured prototype is also very low for all ports ($<0.05$ in band), with the correlation between port 1 and 2 highest in band. 

Overall, the measured results compare reasonably closely with simulation results, and experimentally validate the proposed design method using mode-based synthesis method and 3D printing method.


\section{Conclusion}
\label{section:conclusion}
This paper proposes a systematic feed-independent shape synthesis technique for non-canonical MIMO DRAs based on characteristic mode theory and binary genetic algorithm. This method allows automated synthesis of a MIMO DRAs that support a number of resonant modes less than or equal to those supported by the bounding volume as described by the bounds on MIMO DRAs~\cite{yang2021fundamental}.

While this optimization method is generally applicable to any volume and a variety of operating characteristics and objective criteria, two examples are shown here.  Two 3-port DRAs operating at 2.45 GHz are synthesized from different initial volumes, and one is selected for fabrication using 3D ceramic printing.  Experimental characterization verified the simulated performance and design approach. Good matching ($>10$ dB) and isolation ($>20$ dB) along with low pattern correlation is achieved at the frequency of interest.

\section{Acknowledgement}
This material is based upon work supported in part by the National Science Foundation under Grant No. 2138741. 
The authors also would like to thank Clem Shelton for helping with the DRA's S parameter measurement.

\bibliographystyle{IEEEtran}
\bibliography{bib_May2020}

\begin{thebibliography}{10}
\providecommand{\url}[1]{#1}
\csname url@samestyle\endcsname
\providecommand{\newblock}{\relax}
\providecommand{\bibinfo}[2]{#2}
\providecommand{\BIBentrySTDinterwordspacing}{\spaceskip=0pt\relax}
\providecommand{\BIBentryALTinterwordstretchfactor}{4}
\providecommand{\BIBentryALTinterwordspacing}{\spaceskip=\fontdimen2\font plus
\BIBentryALTinterwordstretchfactor\fontdimen3\font minus \fontdimen4\font\relax}
\providecommand{\BIBforeignlanguage}[2]{{%
\expandafter\ifx\csname l@#1\endcsname\relax
\typeout{** WARNING: IEEEtran.bst: No hyphenation pattern has been}%
\typeout{** loaded for the language `#1'. Using the pattern for}%
\typeout{** the default language instead.}%
\else
\language=\csname l@#1\endcsname
\fi
#2}}
\providecommand{\BIBdecl}{\relax}
\BIBdecl

\bibitem{long1983}
S.~{Long}, M.~{McAllister}, and {Liang Shen}, ``The resonant cylindrical dielectric cavity antenna,'' \emph{IEEE Transactions on Antennas and Propagation}, vol.~31, no.~3, pp. 406--412, May 1983.

\bibitem{lai2008comparison}
Q.~Lai, G.~Almpanis, C.~Fumeaux, H.~Benedickter, and R.~Vahldieck, ``Comparison of the radiation efficiency for the dielectric resonator antenna and the microstrip antenna at ka band,'' \emph{IEEE Transactions on Antennas and Propagation}, vol.~56, no.~11, pp. 3589--3592, 2008.

\bibitem{chair2004wideband}
R.~Chair, A.~Kishk, K.~Lee, and C.~Smith, ``Wideband flipped staired pyramid dielectric resonator antennas,'' \emph{Electronics Letters}, vol.~40, no.~10, pp. 581--582, 2004.

\bibitem{huang2007compact}
W.~Huang and A.~Kishk, ``Compact wideband multi-layer cylindrical dielectric resonator antennas,'' \emph{IET Microwaves, Antennas \& Propagation}, vol.~1, no.~5, pp. 998--1005, 2007.

\bibitem{yan2011design}
J.-B. Yan and J.~T. Bernhard, ``Design of a mimo dielectric resonator antenna for lte femtocell base stations,'' \emph{IEEE Transactions on Antennas and Propagation}, vol.~60, no.~2, pp. 438--444, 2011.

\bibitem{abdalrazik2017three}
A.~Abdalrazik, A.~S.~A. El-Hameed, and A.~B. Abdel-Rahman, ``A three-port mimo dielectric resonator antenna using decoupled modes,'' \emph{IEEE Antennas and Wireless Propagation Letters}, vol.~16, pp. 3104--3107, 2017.

\bibitem{chow1995cylindrical}
K.~Chow, K.~Leung, K.~Luk, and E.~Yung, ``Cylindrical dielectric resonator antenna array,'' \emph{Electronics Letters}, vol.~31, no.~18, pp. 1536--1537, 1995.

\bibitem{su2016linearly}
M.~Su, L.~Yuan, and Y.~Liu, ``A linearly polarized radial-line dielectric resonator antenna array,'' \emph{IEEE Antennas and Wireless Propagation Letters}, vol.~16, pp. 788--791, 2016.

\bibitem{zhang2019mimo}
Y.~Zhang, J.-Y. Deng, M.-J. Li, D.~Sun, and L.-X. Guo, ``A mimo dielectric resonator antenna with improved isolation for 5g mm-wave applications,'' \emph{IEEE Antennas and Wireless Propagation Letters}, vol.~18, no.~4, pp. 747--751, 2019.

\bibitem{nor2016rectangular}
N.~M. Nor, M.~H. Jamaluddin, M.~R. Kamarudin, and M.~Khalily, ``Rectangular dielectric resonator antenna array for 28 ghz applications,'' \emph{Progress In Electromagnetics Research}, vol.~63, pp. 53--61, 2016.

\bibitem{Leung1993Theory}
K.~W. {Leung}, K.~M. {Luk}, K.~Y.~A. {Lai}, and D.~{Lin}, ``Theory and experiment of a coaxial probe fed hemispherical dielectric resonator antenna,'' \emph{IEEE Transactions on Antennas and Propagation}, vol.~41, no.~10, pp. 1390--1398, 1993.

\bibitem{fang2014theory}
X.~S. Fang, K.~W. Leung, and K.~M. Luk, ``Theory and experiment of three-port polarization-diversity cylindrical dielectric resonator antenna,'' \emph{IEEE Transactions on Antennas and Propagation}, vol.~62, no.~10, pp. 4945--4951, 2014.

\bibitem{roper2014additive}
D.~A. Roper, B.~L. Good, R.~McCauley, S.~Yarlagadda, J.~Smith, A.~Good, P.~Pa, and M.~S. Mirotznik, ``Additive manufacturing of graded dielectrics,'' \emph{Smart Materials and Structures}, vol.~23, no.~4, p. 045029, 2014.

\bibitem{kadvera2022wide}
P.~Kad{\v{e}}ra, J.~S{\'a}nchez-Pastor, H.~Eskandari, T.~Tyc, M.~Sakaki, M.~SCH{\"U}{\ss}LER, R.~Jakoby, N.~Benson, A.~Jim{\'e}nez-S{\'a}ez, and J.~L{\'a}{\v{c}}{\'\i}k, ``Wide-angle ceramic retroreflective luneburg lens based on quasi-conformal transformation optics for mm-wave indoor localization,'' \emph{IEEE Access}, vol.~10, pp. 41\,097--41\,111, 2022.

\bibitem{lou2020design}
Y.-H. Lou, Y.-X. Zhu, G.-F. Fan, W.~Lei, W.-Z. Lu, and X.-C. Wang, ``Design of ku-band flat luneburg lens using ceramic 3-d printing,'' \emph{IEEE Antennas and Wireless Propagation Letters}, vol.~20, no.~2, pp. 234--238, 2020.

\bibitem{oh2019microwave}
Y.~Oh, V.~Bharambe, B.~Mummareddy, J.~Martin, J.~McKnight, M.~A. Abraham, J.~M. Walker, K.~Rogers, B.~Conner, P.~Cortes \emph{et~al.}, ``Microwave dielectric properties of zirconia fabricated using nanoparticle jetting™,'' \emph{Additive Manufacturing}, vol.~27, pp. 586--594, 2019.

\bibitem{Oh2023}
Y.~Oh, N.~Kordsmeier, H.~Askari, and J.~J. Adams, ``Low profile grin lenses with integrated matching using 3-d printed ceramic,'' \emph{IEEE Open Journal of Antennas and Propagation}, vol.~4, pp. 12--22, 2023.

\bibitem{xia20193D}
Z.-X. Xia, K.~W. Leung, and K.~Lu, ``3-d-printed wideband multi-ring dielectric resonator antenna,'' \emph{IEEE Antennas and Wireless Propagation Letters}, vol.~18, no.~10, pp. 2110--2114, 2019.

\bibitem{trinh2016wideband}
S.~Trinh-Van, Y.~Yang, K.-Y. Lee, and K.~C. Hwang, ``A wideband circularly polarized pixelated dielectric resonator antenna,'' \emph{Sensors}, vol.~16, no.~9, p. 1349, 2016.

\bibitem{alroughani2020shape}
H.~{Alroughani} and D.~A. {McNamara}, ``The shape synthesis of dielectric resonator antennas,'' \emph{IEEE Transactions on Antennas and Propagation}, vol.~68, no.~8, pp. 5766--5777, 2020.

\bibitem{yang2016systematic}
B.~Yang and J.~J. Adams, ``Systematic shape optimization of symmetric {MIMO} antennas using characteristic modes,'' \emph{IEEE Trans. Antennas Propag.}, vol.~64, no.~7, pp. 2668--2678, 2016.

\bibitem{yang2019shape}
B.~Yang, J.~Zhou, and J.~J. Adams, ``A shape-first, feed-next design approach for compact planar mimo antennas,'' \emph{Progress In Electromagnetics Research}, vol.~77, pp. 157--165, 2019.

\bibitem{harrington1971theory}
R.~Harrington and J.~Mautz, ``Theory of characteristic modes for conducting bodies,'' \emph{IEEE transactions on antennas and propagation}, vol.~19, no.~5, pp. 622--628, 1971.

\bibitem{ethier2014antenna}
J.~L. Ethier and D.~A. McNamara, ``Antenna shape synthesis without prior specification of the feedpoint locations,'' \emph{IEEE Transactions on Antennas and Propagation}, vol.~62, no.~10, pp. 4919--4934, 2014.

\bibitem{filtering_antenna_synthesis}
K.~Li and Y.~Shi, ``Filtering antenna synthesis based on characteristic mode theory,'' \emph{IEEE Transactions on Antennas and Propagation}, vol.~70, no.~5, pp. 3308--3319, 2022.

\bibitem{ESA_synthesis}
R.~Li and G.~Wei, ``Electrically small antenna shape synthesis that approaches the maximum achievable directivity,'' \emph{IEEE Antennas and Wireless Propagation Letters}, vol.~21, no.~6, pp. 1198--1202, 2022.

\bibitem{alroughani2013appraisal}
H.~Alroughani, ``An appraisal of the characteristic modes of composite objects,'' Ph.D. dissertation, University of Ottawa, 2013.

\bibitem{yang2017quality}
B.~Yang and J.~J. Adams, ``Quality factor calculations for the characteristic modes of dielectric resonator antennas,'' in \emph{Radio Science Meeting (USNC-URSI NRSM), 2017 United States National Committee of URSI National}.\hskip 1em plus 0.5em minus 0.4em\relax IEEE, 2017, pp. 1--2.

\bibitem{yang2021fundamental}
B.~Yang, J.~Kim, and J.~J. Adams, ``Fundamental limits on substructure dielectric resonator antennas,'' \emph{IEEE Open Journal of Antennas and Propagation}, vol.~3, pp. 59--68, 2021.

\bibitem{blanch2003exact}
S.~Blanch, J.~Romeu, and I.~Corbella, ``Exact representation of antenna system diversity performance from input parameter description,'' \emph{Electronics letters}, vol.~39, no.~9, pp. 705--707, 2003.

\end{thebibliography}

\end{document}